\documentclass[aip,pof,amsmath,amssymb,preprint]{revtex4-1}

\usepackage{tabularx}
\newcolumntype{C}[1]{>{\centering\arraybackslash}p{#1}}
\usepackage{textcase}
\usepackage{graphicx}
\usepackage{units}
\usepackage{siunitx}
\usepackage{color}

\usepackage{multirow}
\usepackage{tikz}
\usetikzlibrary{shapes,arrows}
\usepackage{pgfplots}
\usepgfplotslibrary{groupplots}
\usetikzlibrary{spy,decorations.fractals,shapes.misc}
\usetikzlibrary{decorations.markings}
\usepgfplotslibrary{colormaps}

\usepackage[caption=false]{subfig}
\usepackage{epstopdf}
    
\begin{document}

\title[A Particle-based ESBGK Method for Diatomic Molecules]{Extending the Particle ESBGK Method to Diatomic Molecules including Quantized Vibrational Energies} 

\author{M. Pfeiffer}
 \email{mpfeiffer@irs.uni-stuttgart.de}
\affiliation{%
Institute of Space Systems, University of Stuttgart, Pfaffenwaldring 29, D-70569
   Stuttgart, Germany
}%

\date{\today}

\begin{abstract}
The particle-based ellipsoidal statistical Bhatnagar-Gross-Krook (ESBGK) model 
is extended to diatomic molecules and compared with the Direct Simulation Monte Carlo (DSMC) method. 
For this an efficient method is developed that optionally allows the handling of quantized vibrational energies.
The proposed method is verified with a gas in an adiabatic box relaxing from a non-equilibrium state to an equilibrium.
It is shown that the analytical Landau-Teller expression as well as DSMC results agree very well with the new method. 
Furthermore, the method is compared with DSMC results and experimental measurements of a hypersonic
flow around a 70$^\circ$ blunted cone. 
It is shown that the ellipsoidal statistical BGK compares very well with the DSMC results while saving up to a factor of $\approx 35.8$ CPU time for this low Knudsen
number case.
\end{abstract}


\maketitle
\section{Introduction}
Simulations of non-equilibrium gas flows are still challenging especially if the simulation region includes dense and rarefied gas regions. The situation becomes even more complex for a molecular
gas flow. In this case, non-equilibrium effects can also affect the inner energies of the molecules. CFD methods based on the Navier-Stokes equations cover a wide range of near equilibrium flows that 
are important for many practical applications. Nevertheless, the assumptions of the Navier-Stokes equations become invalid for rarefied non-equilibrium flows. Another approach of flow field simulation
is the Direct Simulation Monte Carlo (DSMC) method. In this method, discrete particle collisions are used to mimic the convective and collision molecular process\cite{Bird1994}. Therefore, DSMC is able to 
hanlde non-equilibrium effects but becomes very expensive for small Knudsen number flows due to the fact that molecular events must be resolved in space and time within the mean free path and collision frequency, respectively.

The gap between the applicable flow regimes of both methods can be closed with different approaches. A short overview of these methods including advantages and disadvantages of these methods is given in Mirza et al.\cite{MIRZA2017269}. The main focus in this paper is on the particle-based statistical Bhatnagar-Gross-Krook (BGK) method. This method is already used and coupled to DSMC in different applications like nozzle flow expansion \cite[]{burt2006evaluation}, micro channel flows\cite[]{titov2008analysis} or hypersonic shocks \cite[]{tumuklu2016particle,Pfeiffer2018}. Recently, an efficient method to handle arbitrary target distribution functions in the BGK context was presented\cite{Pfeiffer2018}. However, it was shown that the energy conservation scheme becomes very important to produce the correct heat flux vectors, especially for non-symmteric distribution functions, e.g. resulting from the Shakhov model\cite{shakhov1968generalization}. It was additionally shown that the ellipsoidal statistical BGK (ESBGK) 
model\cite{holway1966new} is very robust and produces good results concerning heat flux and shock structures.

In this paper, a relaxation model of internal energies for the ESBGK method will be presented. This work is based on the works of several authors\cite{gallis2011investigation,tumuklu2016particle,burt2006evaluation} but will allow the handling of quantised vibrational
energies as typically used in the DSMC context. Furthermore, the energy conservation scheme is adapted, so that the relaxation process can also occur if only one particle is involved as opposed to the
method proposed in several publications\cite{tumuklu2016particle, burt2006evaluation}. First, the theory of the ESBGK model as well as internal energies are shortly discussed. Then, the implementation is described and subsequently verified by means of simple reservoir simulations, where it is also compared to the DSMC method. Finally, the method is validated with a hypersonic flow around
a 70$^\circ$ blunted cone including a shock structure.

\section{Theory}
The Boltzmann equation describes the behaviour of gas with the corresponding distribution function 
$f=f(\mathbf x, \mathbf v, t)$ at position $\mathbf x$ and velocity $\mathbf v$
\begin{equation}
\frac{\partial f}{\partial t} + \mathbf v \frac{\partial f}{\partial \mathbf x} = \left.\frac{\delta f}{\delta t}\right|_{Coll}.
\end{equation}
In this equation, external forces are neglected. Furthermore, $\left.\delta f/\delta t\right|_{Coll}$ is the 
collision term, which can be described by the Boltzmann collision integral
\begin{equation}
\left.\frac{\partial f}{\partial t}\right|_{Coll}=\int_{\mathbb{R}^3}\int_{S^2}\mathcal B
\left[f(\mathbf v')f(\mathbf v_*')-f(\mathbf v)f(\mathbf v_*)\right]d\mathbf n d\mathbf v_*.
\end{equation}
Here, $S^2\subset\mathbb{R}^3$ is the unit sphere, $\mathbf n$ is the unit vector of the scattered velocities, $\mathcal B$ is the collision
kernel and the superscript $'$ denotes the post collision velocities. The multiple integration of this collision term makes is difficult to compute. 

\subsection{ESBGK Model}

The ESBGK model approximates the collision term to a simple relaxation form, where the distribution function relaxes towards a target distribution function $f^{ES}$ with a 
certain relaxation frequency $\nu$:
\begin{equation}
\left.\frac{\partial f}{\partial t}\right|_{Coll}=\nu\left(f^{ES}-f\right).
\label{eq:bgkmain}
\end{equation}
The target velocity distribution function $f^{ES}$ is given by 
\begin{equation}
f^{ES}=\frac{n}{\sqrt{\det \mathcal A}} \left(\frac{m}{2\pi k_B T}\right)^{3/2} \exp\left[-\frac{m\mathbf c^T \mathcal A^{-1} \mathbf c}{2k_B T}\right]
\label{eq:esbgkdist}
\end{equation}  
with the anisotropic matrix 
\begin{equation}
\mathcal A = \mathcal I - \frac{1-Pr}{Pr}\left(\frac{3\mathcal P}{\mathrm{Tr}\left[\mathcal P\right]}-\mathcal I\right).
\end{equation}
The anisotropic matrix $\mathcal A$ consists of the identity matrix $\mathcal I$ and the pressure tensor $\mathcal P$,
\begin{equation}
\mathcal P = \int \mathbf c \mathbf c^T f\,d\mathbf v,
\end{equation}
which are both symmetric. 
Additionally, $n$ is the particle density, $m$ the particle mass, $T$ the temperature and $\mathbf c=\mathbf v -\mathbf u$  the thermal particle velocity determined from the 
particle velocity $\mathbf v$ and the average flow velocity $\mathbf u$ \cite{bhatnagar1954model}. 
The ESBGK model reproduces the Maxwellian distribution in the equilibrium state as well as the correct moments
of the Boltzmann equation. Furthermore, Andries et al.\cite{andries2000gaussian,andries2001bgk} have shown that it
fulfills the H-theorem. In the ESBGK model, the viscosity and the thermal conductivity are defined as 
\begin{equation}
\mu=\frac{nk_BT}{\nu}Pr \qquad\qquad K=\frac{c_Pnk_BT}{\nu}
\end{equation}
with the specific heat constant 
$c_P=5k_B/2m$.
Due to the fact that the viscosity depends on the Prandtl number, it is possible to reproduce the viscosity and thermal
conductivity at the same time. Thus, the introduction of the Prandtl number as an additional parameter resolves the Prandtl number
problem of the standard BGK model.
The Prandtl number
of molecules depends on inner degrees of freedom:
\begin{equation} 
Pr = \frac{2(5+\xi_R+\xi_V)}{15+2(\xi_V+\xi_R)}
\end{equation}
with the rotational and vibrational degrees of freedom $\xi_R$, $\xi_V$, respectively.

As proposed by Gallis and Torczynski\cite{gallis2011investigation}, a symmetric transformation matrix $\mathcal S$ with $\mathcal A = \mathcal S \mathcal S$ can be defined. 
Furthermore, a normalized thermal velocity vector $\mathbf C$ is defined as such that $\mathbf c= \mathcal S \mathbf C$. 
Using these definitions, the argument of the exponential function in Eq. \eqref{eq:esbgkdist} becomes
\begin{equation}
\mathbf c^T \mathcal A^{-1} \mathbf c = (\mathcal S\mathbf C)^T \mathcal S^{-1} \mathcal S^{-1} \mathcal S\mathbf C = \mathbf C^T \mathbf C  
\label{eq:smat}
\end{equation}
using $(\mathcal S\mathbf C)^T=\mathbf C^T \mathcal S^T=\mathbf C^T \mathcal S$ due to the fact that $\mathcal S$ is symmetric. Consequently,
$\mathcal S$ can transform a vector $\mathbf C$ sampled from a Maxwellian distribution to a vector $\mathbf c$ sampled from Eq.
\eqref{eq:esbgkdist}.

To determine the correct relaxation frequency $\nu$, the well known exponential ansatz of the viscosity $\mu$
\begin{equation}
\mu=\mu_{ref}\left(\frac{T}{T_{ref}}\right)^{\omega_{VHS}}
\end{equation}
is used. Here, $T_{ref}$ is a reference temperature, $\mu_{ref}$ the reference dynamic viscosity at $T_{ref}$ \cite[]{burt2006evaluation}
and $\omega_{VHS}$ is a parameter of the used variable hard sphere model (VHS).  
For a VHS gas the reference dynamic viscosity can be calculated with the VHS reference diameter $d_{ref}$ of the particles:
\begin{equation}
\mu_{ref}=\frac{30\sqrt{mk_BT_{ref}}}{\sqrt{\pi}4(5-2\omega_{VHS})(7-2\omega_{VHS})d_{ref}^2}.
\end{equation}

\subsection{Diatomic Molecules}
Next to the translational energy also the relaxation of internal vibrational and rotational energies must be treated in flows including molecules.
The relaxation of the rotational temperature $T_R$ and the vibrational temperature $T_V$ is typically described with the Landau-Teller 
equation\cite{zhang:2013, Gimelshein2002, pfeiffer2016direct}
\begin{equation}
\frac{d T_i}{dt}=-\nu_i(T_i-T_{Eq,i}),\quad i=R,\,V,
\label{eq:landtel}
\end{equation}
with the corresponding relaxation frequency $\nu_i$ and the equilibrium temperature $T_{Eq,i}$ for the rotational and vibrational energy.
The equilibrium temperature in the Landau-Teller equation is the instantaneous translational cell temperature $T_{Eq,i}=T_{tr}$,
which is calculated by using the equation of the unbiased sample variance as described in Sun and Boyd\cite{sun2005evaluation} 
\begin{equation}
T_{tr}=\frac{m}{3 k_B}\frac{\sum_{i=1}^N \mathbf{c}^2_i}{N-1}.
\label{eq:transtemp}
\end{equation} 
Eq. \eqref{eq:transtemp} results in $\xi_{tr} = 3 (N-1)/N$ effective translational degrees of freedom per particle according to the equipartition theorem.


The rotational temperature $T_R$ of the molecules in a diatomic rigid rotator model can be calculated using 
\begin{equation}
T_R=\frac{2}{\xi_Rk_B}\frac{\sum_{i=1}^N E_{R,i}}{N}
\end{equation}
with the rotational energy $E_{R,i}$ of particle $i$ and the rotational degrees of freedom $\xi_R=2$.

The vibrational energy is described by the harmonic oscillator model
\begin{equation}
E_V=(j+0.5)k_B\theta_V,
\end{equation}
with the vibrational quantum number $j$ and the characteristic vibrational temperature $\theta_V$.
The analytical solution of the vibrational temperature in this
model is given by 
\begin{equation}
T_V=\frac{\theta_V}{\ln\left[1+1/(\left<E_V\right>/k_B\theta_V -0.5)\right]}, \quad\quad 
\left<E_V\right>=\frac{\sum_{i=1}^N E_{V,i}}{N}.
\end{equation}
The vibrational degrees of freedom are depending on the vibrational temperature and can be calculated with 
\begin{equation}
\xi_V(T_V)=\frac{2\left<E_V\right>}{k_B T_V}\quad\text{or}\quad \xi_V(T_V)=\frac{2\theta_V/T_V}{e^{\theta_V/T_V}-1}.
\end{equation}

The relaxation frequency $\nu_i$ of the inner degrees of freedom depends on the collision frequency of the gas $\nu_{coll}$
\begin{equation}
\nu_{coll}=2d_{ref}^2n \sqrt{\frac{4\pi k_B T_{ref}}{m}}\left(\frac{T_{ref}}{T}\right)^{\omega_{VHS}}
\end{equation} 
according to $\nu_i=\nu_{coll}/Z_i$ with the collision number $Z_i$. Different models exist for the vibrational and rotational collision numbers $Z_i$,
which can be found in Gimelshein et al.\cite{Gimelshein2002,Haas1994}. As a simplification, constant collision numbers are assumed here. However, these constant numbers can easily be replaced with 
more sophisticated models in the proposed method.

\section{Implementation}
The ESBGK particle method is implemented in the PIC-DSMC code PICLas \cite[]{Munz2014} as described in detail in Pfeiffer\cite{Pfeiffer2018}. 

The main concept of the particle ESBGK method especially the energy and momentum conservation, is based on the works of 
\cite{gallis2011investigation,gallis2000application,burt2006evaluation,tumuklu2016particle}. 
Here, particles are moved in a simulation mesh,
collide with boundaries and the microscopic particle properties are sampled to calculate macroscopic values in the same manner as in 
DSMC. But in contrast to the DSMC method, the collision step with binary collisions between the particles is not performed. Instead,
each particle in a cell relaxes with the probability
\begin{equation}
P=1-\exp\left[-\nu \Delta t\right]
\label{eq:bgkrelax}
\end{equation}
according to Eq. \eqref{eq:bgkmain} towards the target distribution. The relaxation frequency $\nu$ is evaluated in each time step for each
cell from the definition of the viscosity of each model. The relaxation frequency directly depends on the cell temperature $T$, which
is calculated from the particle information. 

If a particle is chosen to relax, the new particle velocity is sampled from the target distribution. 
The detailed description of the sampling process for different target distributions (e.g. ESBGK or SBGK) can be found in Pfeiffer\cite{Pfeiffer2018}.
Here, an approach is used with 
an approximation of the transformation matrix $\mathcal S$ of eq. \eqref{eq:smat} as described 
in previous studies \cite{gallis2011investigation, burt2006evaluation,tumuklu2016particle}
\begin{equation}
\mathcal S_{ij} = \delta_{ij}-\frac{1-Pr}{2Pr}\left[\frac{m}{k_B T}\frac{N}{N-1}\left(\mathcal P_{ij}-{\hat{c}_i}{\hat{c}_j}\right)-\delta_{ij}\right]
\label{eq:Sconvert}
\end{equation}
with 
\begin{equation}
\mathbf{\hat{c}} = \int \mathbf c f\,d\mathbf v.
\end{equation}

\subsection{Relaxation of Internal Energies}
To incorporate the internal energies in the algorithm, the Landau-Teller equation \eqref{eq:landtel} is used in the same manner as the
BGK collision term of Eq. \eqref{eq:bgkmain}. This means that each molecule relaxes the internal degree of freedom $i$ to the equilibrium temperature Eq.
\eqref{eq:transtemp} with the probability
\begin{equation}
P_i=1-\exp\left[-\nu_i \Delta t\right],\quad i=R,\,V,
\label{eq:innerrelprob}
\end{equation}
irrespective of whether the particle is already chosen to relax according to Eq. \eqref{eq:bgkrelax}.
Whether or not this procedure reproduces the temperature according to the Landau-Teller equation depends on the used energy conservation scheme
as described in Burt and Boyd\cite{burt2006evaluation} and Tumuklu et al.\cite{tumuklu2016particle}.
The energy conservation is done by involving all $N$ particles in a cell instead of only the relaxing particles. 
This scheme has shown to be more accurate in the case of small particle numbers per cell in different test cases\cite{Pfeiffer2018}. 
The conditions for fulfilling energy and momentum conservation assuming only one species with mass $m$ are
\begin{eqnarray}
m\sum_{i=1}^N \mathbf v^*_i &&= m\sum_{i=1}^N \mathbf v_i \\
\sum_{i=1}^N \frac{m}{2}\left(\mathbf v^*_i\right)^2 +\sum_{i=1}^{N_R} E^*_{R,i} + \sum_{i=1}^{N_V} E^*_{V,i} &&= \sum_{i=1}^N \frac{m}{2}\mathbf v_i^2 +\sum_{i=1}^{N_R} E_{R,i} + \sum_{i=1}^{N_V} E_{V,i}.\label{eq:energcon}
\end{eqnarray}
Here, the superscript $^*$ denotes the values after the energy conservation process and $N_R$ as well as $N_V$ are the numbers of the rotational and vibrational relaxing molecules 
as the result of Eq. \eqref{eq:innerrelprob}, respectively. 
Eq. \eqref{eq:energcon} yields the following equilibrium temperature after the energy conservation process:
\begin{equation}
T^*_{Eq}=\frac{3(N-1)T_{tr}+2N_RT_R+\xi_V(T_V)N_VT_V}{3(N-1)+2N+\xi_V(T^*_{Eq})N_V}.
\end{equation}
Unfortunately, this equilibrium does not necessarily fulfill the required condition of the Landau-Teller relaxation in Eq. \eqref{eq:landtel}.
To overcome this problem, the relaxation probability of the internal energies is corrected with a parameter $\beta_i$ as suggested by Burt \cite{burt2006evaluation}:
\begin{equation}
P_i=1-\exp\left[-\beta_i\nu_i \Delta t\right],\quad i=R,\,V.
\label{eq:correctPinner}
\end{equation}
To fulfill the Landau-Teller relaxation form, $\beta_i$ will be chosen to
\begin{equation}
\frac{d T_i}{dt}=-\nu_i(T_i-T_{tr})=-\beta_i\nu_i(T_i-T^*_{Eq})\quad \Rightarrow \quad \beta_i = \frac{T_i-T_{tr}}{T_i-T^*_{Eq}}.
\end{equation}
The solution of this equation system is solved numerically, due to the fact that $T^*_{Eq}$ directly depends on $\beta_R$ and $\beta_V$.
For this purpose, the following system is solved 
\begin{eqnarray}
T^{*,n+1}_{Eq}&=&\frac{3(N-1)T_{tr}+2P_R(\beta_R^{n})NT_R+\xi_V(T_V)P_V(\beta_V^{n})NT_V}{3(N-1)+2P_R(\beta_R^{n})N+\xi_V(T^{*,n}_{Eq})P_V(\beta_V^{n})N} \label{eq:tequinum}\\
\beta_i^{n+1} &=& \frac{T_i-T_{tr}}{T_i-T^{*,n+1}_{Eq}}
\end{eqnarray}
until $T^{*,n+1}_{Eq}-T^{*,n}_{Eq}<\varepsilon$ with the iteration step $n$ and an accuracy $\varepsilon$. However, this equation system can be solved with 
an arbitrary method.

The new rotational energy $E'_{R,i}$ of molecules that are chosen for a rotational relaxation according to the corrected probability in Eq. \eqref{eq:correctPinner}
is reassigned to a value from a Boltzmann distribution
\begin{equation}
E'_{R,i}=-k_B T^*_{Eq} \ln(R_i)
\end{equation}
with the random numer $R_i$. This energy must be scaled additionally to fulfill the energy conservation.

The same method is also used for the vibrational energy:
\begin{equation}
E'_{V,i}=-\frac{\xi_V(T^*_{Eq})}{2}k_B T^*_{Eq} \ln(R_i).
\end{equation}
If the vibrational energy should be described by discrete quantum numbers, an additional step is necessary in the energy conservation process of $E_{V,i}$.

\subsection{Energy and Momentum Conservation}
\begin{figure}
\centering
\includegraphics{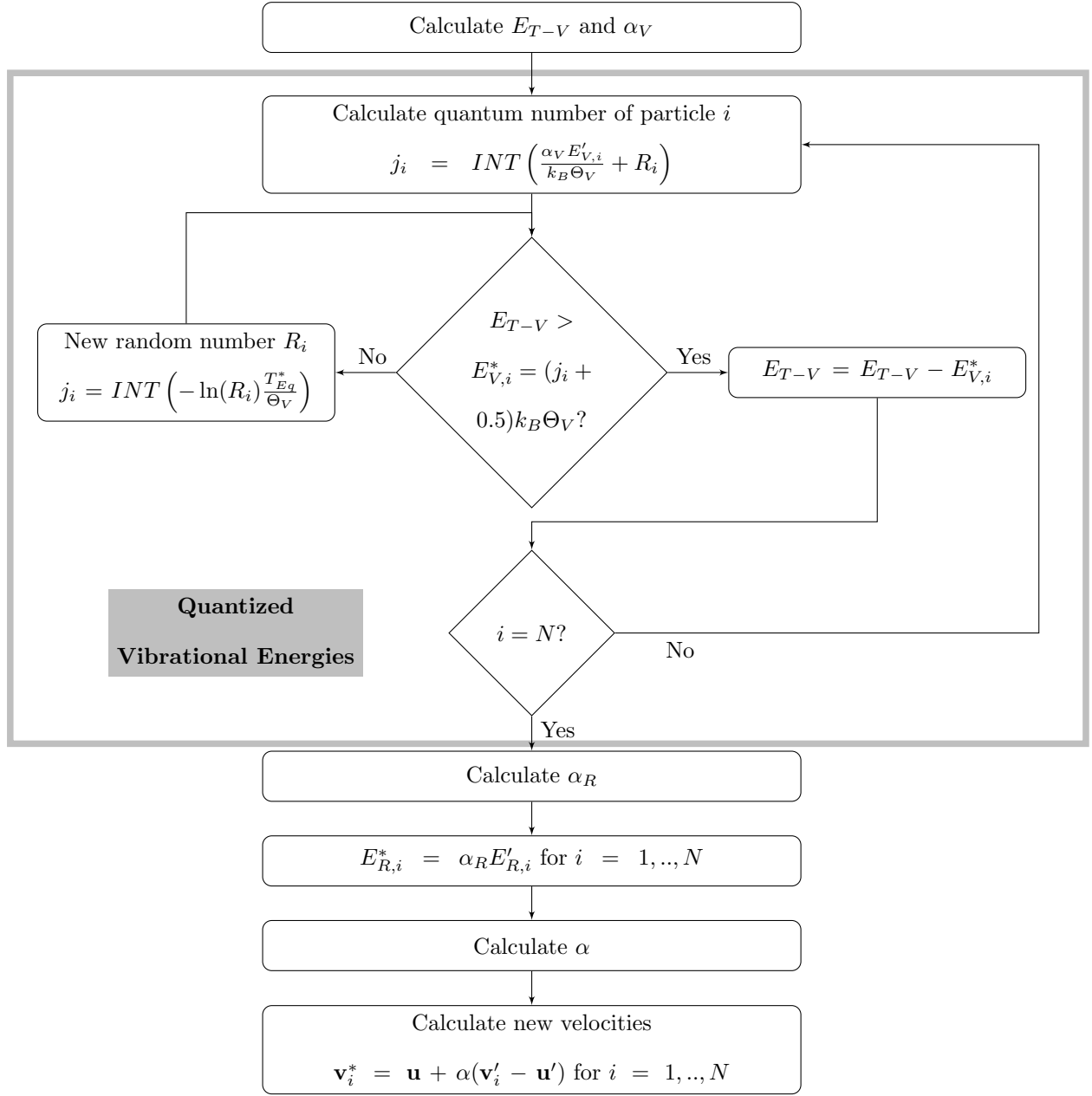}
\caption{Flow chart of energy conservation scheme.}
\label{fig:Energyflowchart}
\end{figure}

The energy conservation process is performed in different steps and illustrated in Fig. \ref{fig:Energyflowchart}. 
First of all, the energy conservation of the vibrational energy is performed. This must be done, if the vibrational energy is described in a quantized way. In the case of continuous vibrational energy, this fixed order is not necessary.
In the used scheme, only translation-vibration ($T-V$) and translation-rotation ($T-R$) relaxation processes are allowed directly. Therefore, the energy $E_{T-V}$
\begin{equation} 
E_{T-V}=\sum_{i=1}^N\frac{m}{2}\mathbf c_i^2 + \sum_{i=1}^{N_V} (E_{V,i}-0.5k_B \Theta_V)
\end{equation}
should be equally distributed over the translational $3(N-1)$ and vibrational $\xi_V(T_{Eq}^*)N_V$ degrees of freedom
to fulfill energy conservation as well as the assumptions of Eq. \eqref{eq:tequinum}.
For this purpose, an $\alpha_V$ is defined with
\begin{equation}
E^*_{V,i}=\alpha_VE'_{V,i} + 0.5 k_B \Theta_V .
\end{equation} 
The equal distribution over the DOFs is reached if
\begin{equation}
\alpha_V=\frac{E_{T-V}}{\sum_{i=1}^{N_V} E'_{V,i}}\left(\frac{\xi_V(T_{Eq}^*)N_V}{\xi_V(T_{Eq}^*)N_V+3(N-1)}\right).
\label{eq:alphaV}
\end{equation} 
For continuous vibrational energies Eq. \eqref{eq:alphaV} is the final step, however, for quantized
vibrational energies, further steps are necessary.

In the following, the quantized energy states for each particle are determined consecutively.
The term $\alpha_VE'_{V,i}$ is reformulated to a quantum number using the random number $R_i$.
\begin{equation}
j_i=INT\left(\frac{\alpha_VE'_{V,i}}{k_B \Theta_V}+R_i\right).
\end{equation}
With this quantum number it is checked whether the condition 
\begin{equation}
E_{T-V}>E^*_{V,i}=(j_i+0.5)k_B \Theta_V
\end{equation} 
is fulfilled. If this is the case, $E_{T-V}$ is updated with $E_{T-V}=E_{T-V}-E^*_{V,i}$
and the next particle is processed.
Otherwise, a new quantum number is calculated with the new random number $R_i$:
\begin{equation}
j_i=INT\left(-\ln(R_i)\frac{T^*_{Eq}}{\Theta_V}\right)
\end{equation}
until the condition $E_{T-V}>E^*_{V,i}$ is fulfilled.
Consequently, also the following particles in this algorithm can have a vibrational energy greater than zero. Note that for the following part of the energy conservation scheme, 
all $E^*_{V,i}$ must be subtracted from $E_{T-V}$ also in the continuous vibrational energy case.

The energy conservation of the rotational and translational energies is achieved analogously to the vibrational energy. This means again that the energy $E_{T-R}$
\begin{equation} 
E_{T-R}=E_{T-V} + \sum_{i=1}^{N_R} E_{R,i}
\end{equation}
should be equally distributed over the translational and rotational degrees of freedom.
Note that $E_{T-V}$ includes the translational energy as well as the remaining vibrational energy in the quantized case. Therefore, the total system energy will be conserved with this scheme. The rotational energy is conserved using
\begin{eqnarray}
E^*_{R,i}&=&\alpha_RE'_{R,i} \\
\alpha_R &=& \frac{E_{T-R}}{\sum_{i=1}^{N_R} E'_{R,i}}\left(\frac{2N_R}{2N_R+3(N-1)}\right).
\end{eqnarray}
The momentum and energy conservation for the translational energy is done as described in the publications 
\cite{Pfeiffer2018,gallis2011investigation} for the ESBGK model. Therefore, the final velocities of the particles are
\begin{equation}
\mathbf v_i^*=\mathbf u + \alpha(\mathbf v'_i-\mathbf u')
\label{eq:transenmom}
\end{equation}
whether they relax or not. Here, $\mathbf u=\sum_{i=1}^N \mathbf v_i/N$ is the average flow velocity before the relaxation, 
$\mathbf v'_i$ are the particle velocities after the relaxation but before the energy conservation process and $\mathbf u'=\sum_{i=1}^N \mathbf v'_i/N$. Note that, 
$\mathbf v'_i = \mathbf v_i$ if no relaxation occurs for particle $i$. Due to
\begin{equation}
\sum_{i=1}^N(\mathbf v'_i-\mathbf u')=0, 
\end{equation} 
eq. \eqref{eq:transenmom} ensures momentum conservation. Energy conservation is achieved by choosing $\alpha$ to
\begin{equation}
\alpha = \sqrt{\frac{E_{T-R}}{\sum_{i=1}^N\frac{m}{2}(\mathbf v'_i-\mathbf u')^2}
\left(\frac{3(N-1)}{2N_R+3(N-1)}\right)}.
\end{equation}

\section{Simulation Results}
\subsection{Reservoir simulations}
\begin{figure}
\centering
\subfloat[Rotational relaxation.\label{fig:ltrot}]{\includegraphics{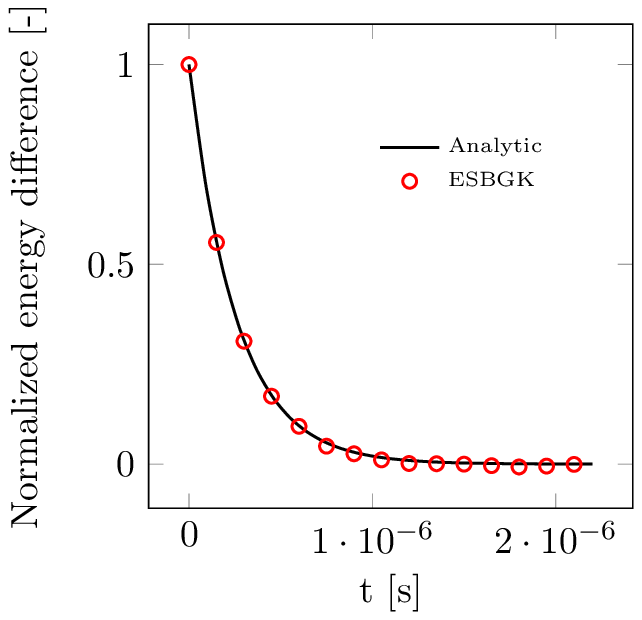}}\quad
\subfloat[Vibrational relaxation.\label{fig:ltvib}]{\includegraphics{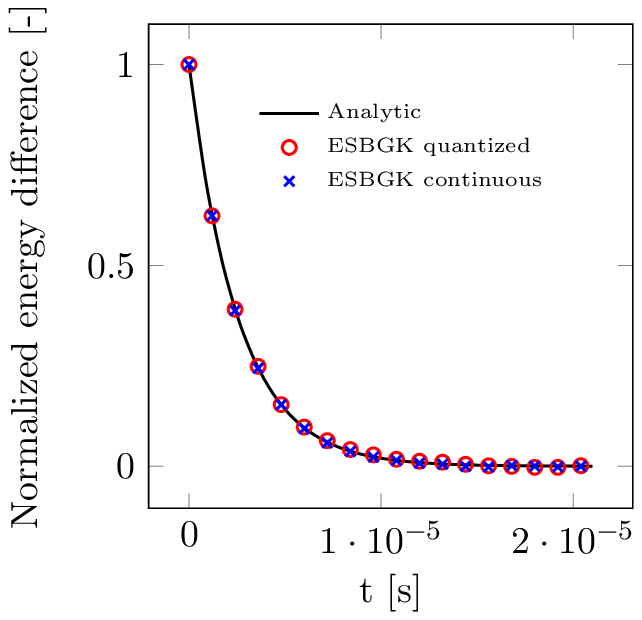}}
\caption{Comparison of BGK simulation results with analytical Landau-Teller solution.}
\label{fig:LT_relax}
\end{figure}

The first verification case is a reservoir simulation in which particles are placed in an adiabatic box. Starting from a thermal non-equilibrium state at $t=0\,\mathrm{s}$, a relaxation process is caused. After a certain amount of time, thermal equilibrium is reached as described by the Landau-Teller equation \eqref{eq:landtel}. If the translational-vibrational and translational-rotational relaxation is considered separately and an isothermal relaxation is assumed ($T_{Eq,i}(t)=T_{Eq,i}(t=\infty)$), it is possible to define an analytical solution of Eq. \eqref{eq:landtel}\cite{Pfeiffer2016b,zhang:2013}:
\begin{equation}
\frac{E_i(\infty)-E_i(t)}{E_i(\infty)-E_i(0)}=e^{-t\nu_i}.
\end{equation}
Here, the Landau-Teller equation is rewritten in the energy form. The verification simulation is done with N$_\mathrm{2}$. The particle density in the simulation is $n\approx2\cdot10^{22}\,\mathrm{m^{-3}}$ ($\num{200000}$ particles with a weighting factor $w=10^7$ in a volume of $V=(4.6\cdot10^{-4})^3\,\mathrm{m^3}$). Initial temperatures of the translational, rotational, and vibrational energies are $T_{\mathrm{tr}}=16000\,\mathrm{K}$, $T_{\mathrm{R}}=8000\,\mathrm{K}$, and $T_{\mathrm{V}}=8000\,\mathrm{K}$, respectively. The characteristic vibrational temperature and species-specific constants for the variable hard sphere (VHS) model are summarized in Table \ref{tab:reservoir_species}. The collision numbers are chosen to 
$Z_R=5$ and $Z_V=50$. For the comparison with the analytical solution, the translational temperature is fixed by 
reassigning the translational energy of the particles according to $T_{\mathrm{tr}}=T_{Eq,i}(t=\infty)=16000\,\mathrm{K}$
after each time step. The rotational and vibrational relaxation is investigated separately by choosing $\nu_V=0$ or $\nu_R=0$, respectively. The results are shown in Fig. \ref{fig:LT_relax}. The analytical solution and the results of the ESBGK method show a very good agreement for the rotational as well as vibrational relaxation. In the vibrational relaxation
case, the quantized and the continuous model are used.

\begin{table}
  \begin{center}
\def~{\hphantom{0}}
  \begin{tabular}{l|c c c c}
   & $\Theta_{V}\,[\mathrm{K}]$ & $\omega_{VHS}$ & $T_{ref}\,[\mathrm{K}]$ & $d_{ref}$ [m] \\ 
  \hline
  N$_\mathrm{2}$ & $3395$ & $0.74$ & $273$ & $4.17\cdot10^{-10}$ 
  \end{tabular}
  \caption{\label{tab:reservoir_species}N$_\mathrm{2}$ species constants.}
  \end{center}
\end{table}

\begin{figure}
\centering
\includegraphics{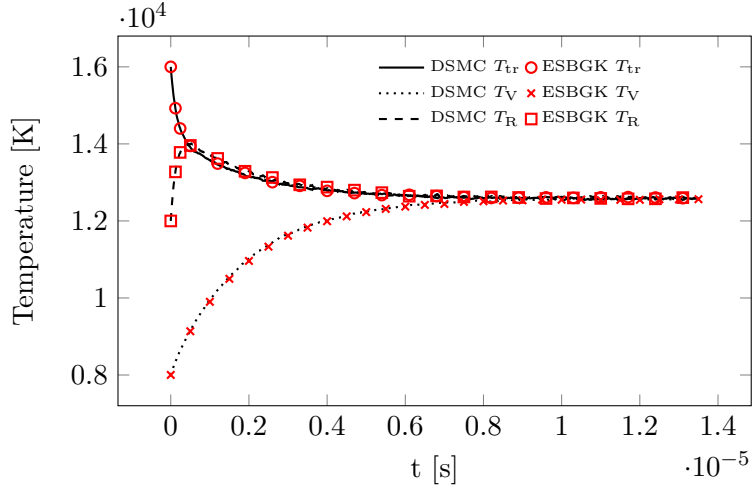}
\caption{Comparison of relaxation process of $T_{\mathrm{tr}}$, $T_{\mathrm{R}}$ and $T_{\mathrm{V}}$ between DSMC
and BGK.}
\label{fig:compdsmcbgk}
\end{figure}

In the second test case, a simultaneous relaxation of the translational, rotational and vibrational temperature is investigated. For this, the reservoir simulation with the conditions described before is used again, only the initial
temperatures are changed to $T_{\mathrm{tr}}=16000\,\mathrm{K}$, $T_{\mathrm{R}}=12000\,\mathrm{K}$, and $T_{\mathrm{V}}=8000\,\mathrm{K}$. The DSMC simulations are done using the prohibiting double relaxation method to reproduce the 
Landau-Teller equation as described in several works\cite{Pfeiffer2016b,zhang:2013}.
The results of the DSMC and the ESBGK simulation using the quantized vibrational model is shown in Fig. \ref{fig:compdsmcbgk}. Excellent agreement is found between the DSMC and the BGK results.  

\subsection{70$^\circ$ Blunted Cone}
\begin{figure}[ht]
\renewcommand*{\arraystretch}{0.6}
\centering
\includegraphics{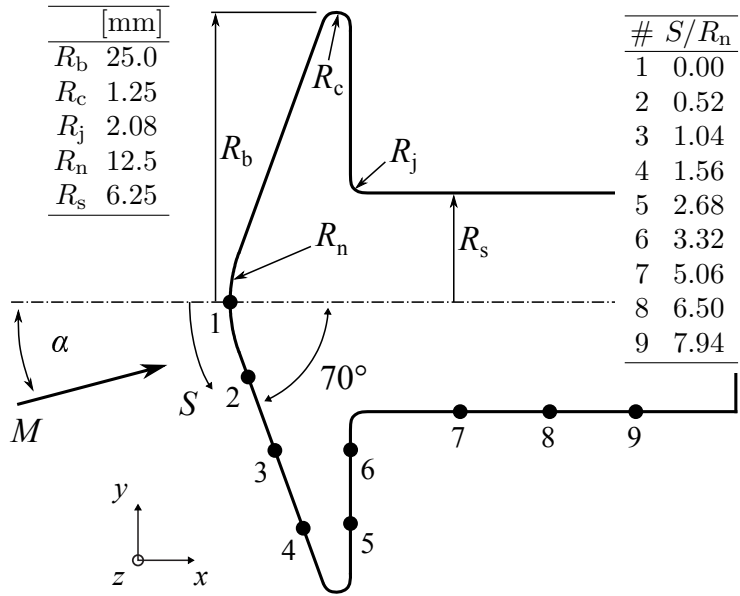}
\caption{$70^{\circ}$ blunted cone geometry and positions of thermocouples.}\label{fig:70degCone}
\end{figure}

The 70$^\circ$ blunted cone described in All{\`{e}}gre et al.\cite{Allegre1997} is chosen to validate the molecular ESBGK implementation, which was also used to validate the DSMC solver\cite{Nizenkov2017}.
The geometry including the position of the thermocouples is shown in Fig \ref{fig:70degCone}. 
Due to the thermocouples, it is possible to compare simulation results with measured heat flux values on the surface.

The simulations were carried out for molecular nitrogen N$_\mathrm{2}$ using the species constants of Table 
\ref{tab:reservoir_species} with the inflow conditions of Table \ref{tab:70cone}.
\begin{table}
  \begin{center}
\def~{\hphantom{0}}
  \begin{tabular}{c c c c c c}
  & $\alpha$ $[^\circ]$ & $\mathbf{v}_\infty$ $\left[\mathrm{ms^{-1}}\right]$ &  $T_{tr,V,R,\infty}$ $[\mathrm{K}]$ &  $n_\infty$ $[\mathrm{m^{-3}}]$ & $Ma$\\ \hline
   \textbf{Case 1} & 0 & $(1502.57, 0.0, 0.0)^T$ & 13.58 & $1.115\cdot10^{21}$ & 20\\
   \textbf{Case 2} & 30 & $(1301.11, 0.0, 751.2)^T$ & 13.58 & $1.115\cdot10^{21}$ & 20 
  \end{tabular}
  \caption{\label{tab:70cone}Inflow conditions of 70$^\circ$ cone test case.}
  \end{center}
\end{table}
The two given inflow velocities are corresponding to angle of attacks of $\alpha=0^\circ$ and $\alpha=30^\circ$. Therefore, full 3D simulations are 
necessary. The Knudsen number of both cases is $Kn\approx0.011$. 
The BGK method has similar requirements
as the CFD method. To resolve the temperature and velocity gradients, a certain number of cells is required. Additionally, a certain number of 
particles per cell is required to represent the moments of the distribution function. Good results using the ESBGK model and the described energy conservation scheme 
are obtained with at least 7 to 10 particles per cell as described in Pfeiffer\cite{Pfeiffer2018}.
The time step can be found using a classic CFL condition with the stream velocity and the speed of sound\cite{MIRZA2017269,PFEIFFER20171}.

\subsubsection*{Case 1 $\alpha=0^\circ$}
\begin{figure}
\centering
\subfloat[Translational Temperature\label{fig:compcase1trans}]{\includegraphics{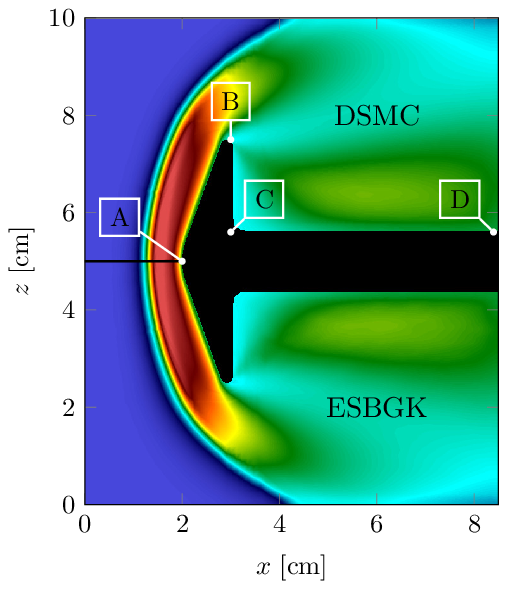}}\quad
\subfloat[Vibrational Temperature]{\includegraphics{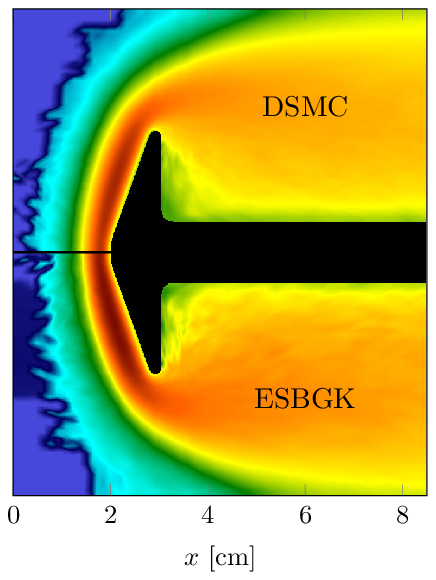}}\quad
\subfloat[Rotational Temperature]{\includegraphics{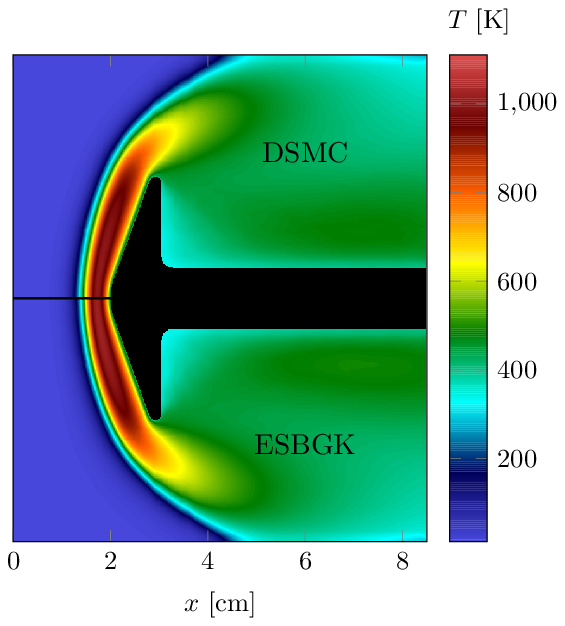}}
\caption{Temperature plots of the flow field using DSMC and ESBGK for case 1 ($\alpha=0^\circ$).\label{fig:compcase1}}
\end{figure}

To resolve the mean free path and the collision frequency in the 3D DSMC simulation, a particle number 
of $N_{DSMC}=4.5\cdot 10^8$ and a time step of $\Delta t_{DSMC}=5\cdot 10^{-8}\,\mathrm{s}$ are necessary. 
A comparison of the translational, rotational and vibrational temperatures in the flow field are shown in Fig. \ref{fig:compcase1}.
The overall qualitative agreement is very good.

\begin{figure}
\centering
\subfloat[Temperature over stagnation stream line.\label{fig:shockcase1}]{\includegraphics{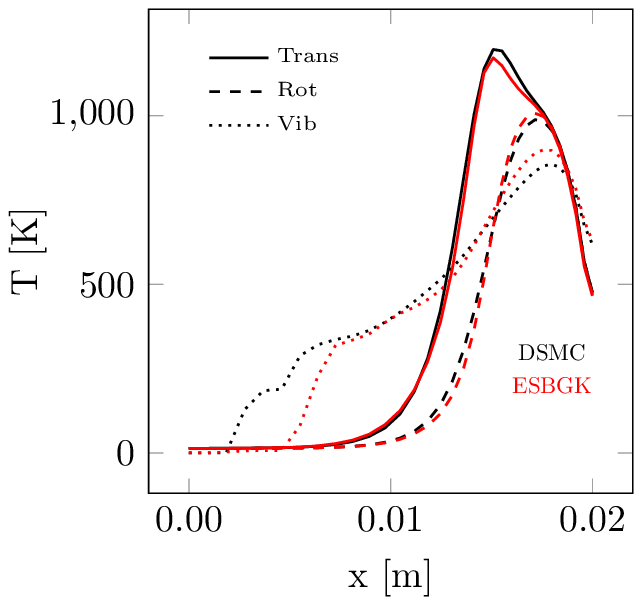}}\quad
\subfloat[Pressure x direction on the surface.\label{fig:forcecase1}]{\includegraphics{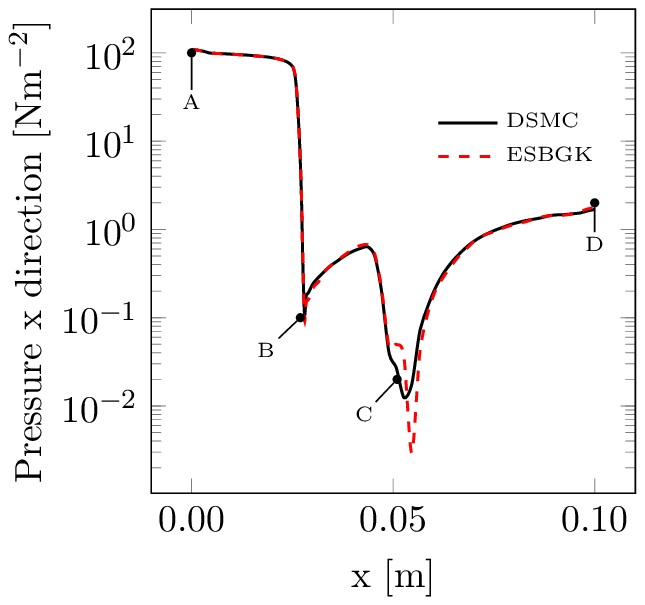}}
\caption{Simulation results of Case 1.}
\end{figure}

The temperature shock profile over the stagnation stream line is shown in detail in Fig. \ref{fig:shockcase1}. The overall agreement of the temperatures
is very good. The biggest difference between the simulations is visible in the inflow area for the vibrational temperature. One reason for this difference is the
statistical noise of the vibrational temperature in this region. The inflow is relatively cold, so that the quantized vibrational temperature is only slightly excited. 
Due to the fact that the ESBGK method needs much less particles ($N_{ESBGK}=N_{DSMC}/16$), the statistical noise is higher in the ESBGK method, which leads to the difference in the 
free stream inflow area.

\begin{figure}
\centering
\includegraphics{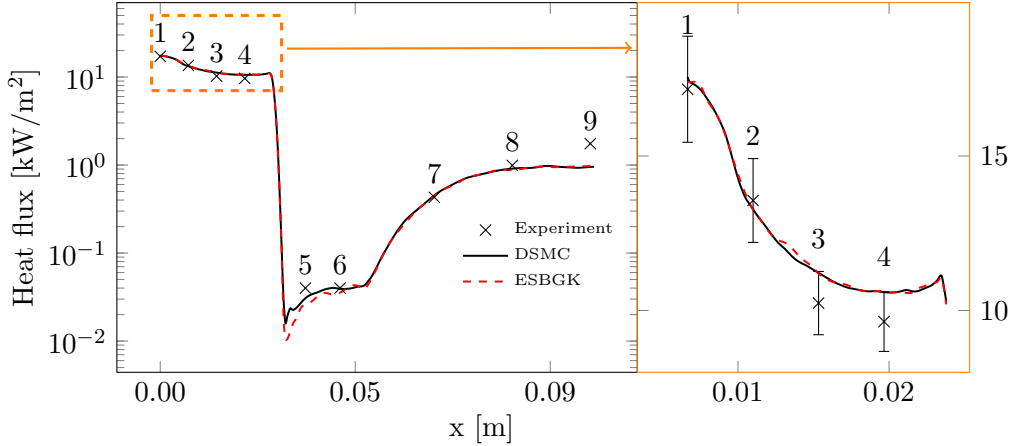}
\caption{Comparison of the heat flux on the surface for case 1.\label{fig:set1heatforcea}}
\end{figure}

The comparison of the heat flux and pressure in x-direction between DSMC and ESBGK are shown in Fig. \ref{fig:set1heatforcea} and \ref{fig:forcecase1}. 
Additionally, the positions and measurements of the thermocouples described in Fig. \ref{fig:70degCone} are shown in Fig. \ref{fig:set1heatforcea}.
The points \{A,B,C,D\} in Fig. \ref{fig:forcecase1} correspond with the points depicted in Fig. \ref{fig:compcase1trans}. 
The DSMC and ESBGK results show a very well agreement. Furthermore, the simulations match the measurements very well.

A comparison of the computational time is shown in Table \ref{tab:cputimeset1}. 
The ESBGK method needs less particles ($N_{ESBGK}=N_{DSMC}/16$) and allows a larger time step $t_{ESBGK}=2t_{DSMC}$. Therefore, the ESBGK model reduces the CPU time by a factor of $\approx 35.8$ 
for this case compared with DSMC.
\begin{table}
  \begin{center}
\def~{\hphantom{0}}
  \begin{tabular}{l | C{2cm} C{2cm} C{3cm} C{4cm}}
   & Particle Number $N$ &  Time step $\Delta t$ [s] &  CPU Time / 100 iterations [s] & CPU Time / $1\cdot 10^{-5}\,\mathrm{s}$ Simulation time [s] \\
  \hline
  DSMC & $4.5\cdot 10^8$& $5\cdot 10^{-8}\,\mathrm{s}$  & 1842 & 3684 \\
  ESBGK & $N_{DSMC}/16$ & $2\Delta t_{DSMC}$ & 103 & 103
  \end{tabular}
  \caption{\label{tab:cputimeset1} Comparison of CPU time between DSMC and ESBGK for Case 1. The CPU time is the time per node with 24 cores on a Intel Xeon CPU E5-2680 v3.}
  \end{center}
\end{table}

\subsubsection*{Case 2 $\alpha=30^\circ$}
\begin{figure}
\centering
\subfloat[Translational Temperature\label{fig:shockline}]{\includegraphics{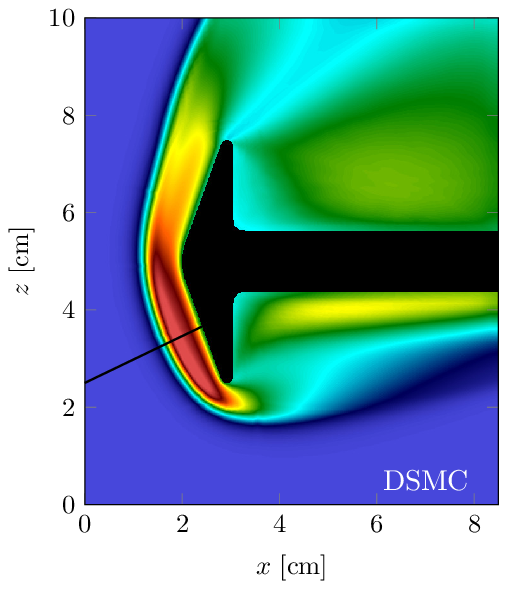}}\quad
\subfloat[Vibrational Temperature]{\includegraphics{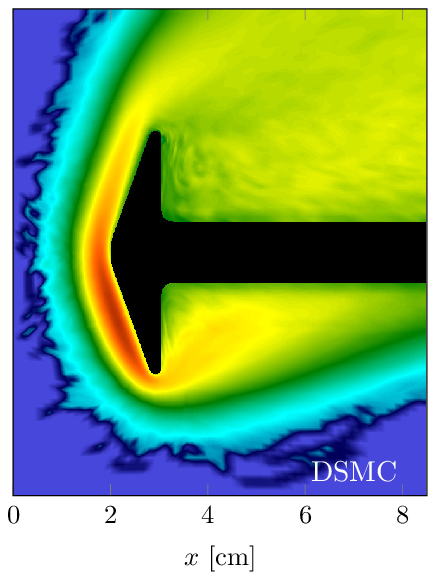}}\quad
\subfloat[Rotational Temperature]{\includegraphics{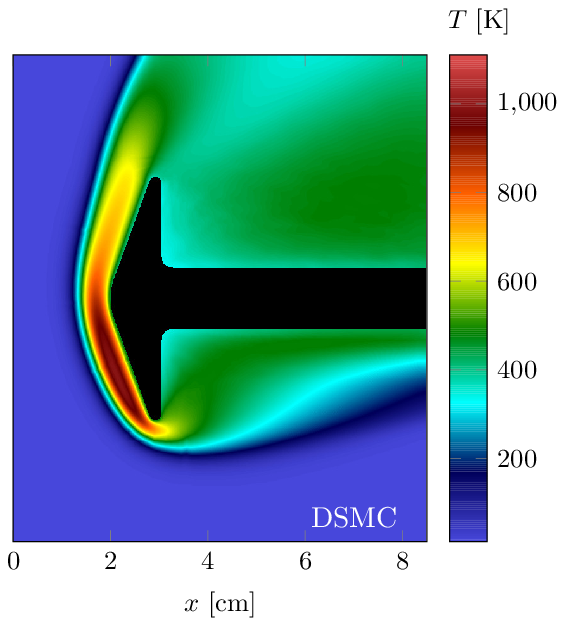}}\\
\subfloat[Translational Temperature]{\includegraphics{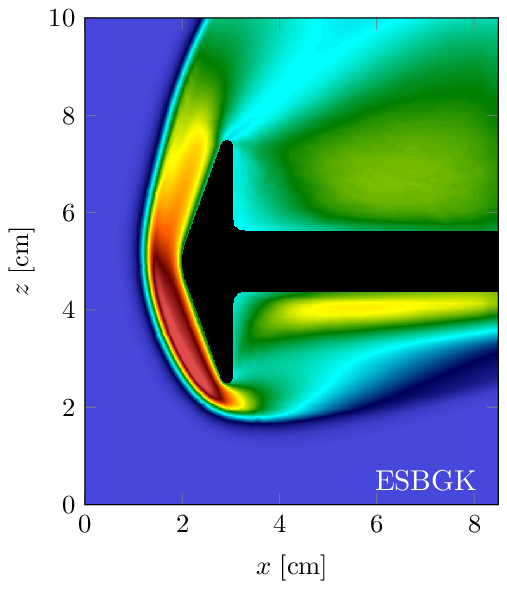}}\quad
\subfloat[Vibrational Temperature]{\includegraphics{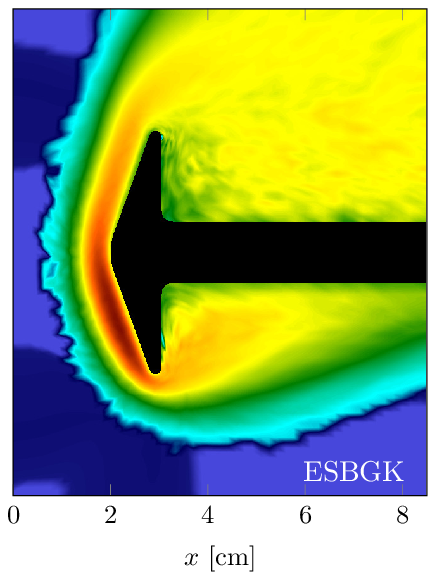}}\quad
\subfloat[Rotational Temperature]{\includegraphics{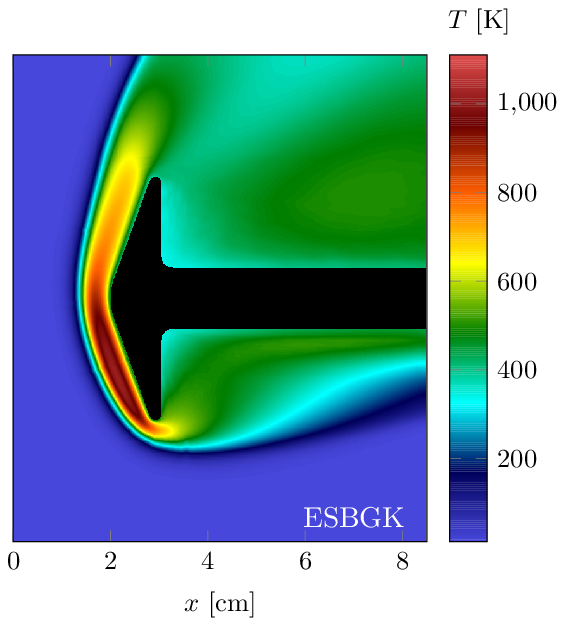}}
\caption{Temperature plots of the flow field using DSMC and ESBGK for case 2 $\alpha=30^\circ$.\label{fig:compcase2}}
\end{figure}

The mean free path and the collision frequency in the 3D DSMC simulation of Case 2 $\alpha=30^\circ$ are resolved using
$N_{DSMC}=4.2\cdot 10^8$ and a time step of $\Delta t_{DSMC}=5\cdot 10^{-8}\,\mathrm{s}$. 
A comparison of the translational, rotational and vibrational temperatures in the flow field is shown in Fig. \ref{fig:compcase2}.
The overall agreement is again very good. The largest differences are visible in the wake region behind the shield. Furthermore, the vibrational temperature is
slightly overestimated in the wake region. However, again much less particles are used in the ESBGK case, leading to greater statistical fluctuations and different results in the wake,
especially for the quantized vibrational temperature.

\begin{figure}
\centering
\subfloat[Temperature over stagnation stream line.\label{fig:shockcase2}]{\includegraphics{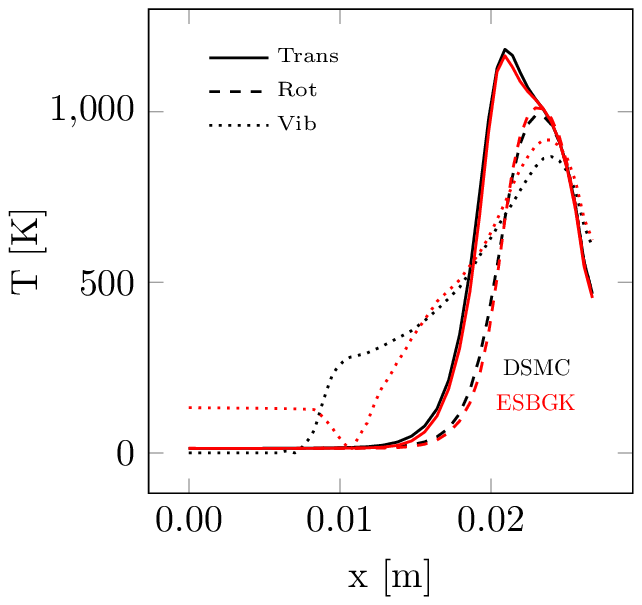}}\quad
\subfloat[Pressure x direction on the surface.\label{fig:forcecase2}]{\includegraphics{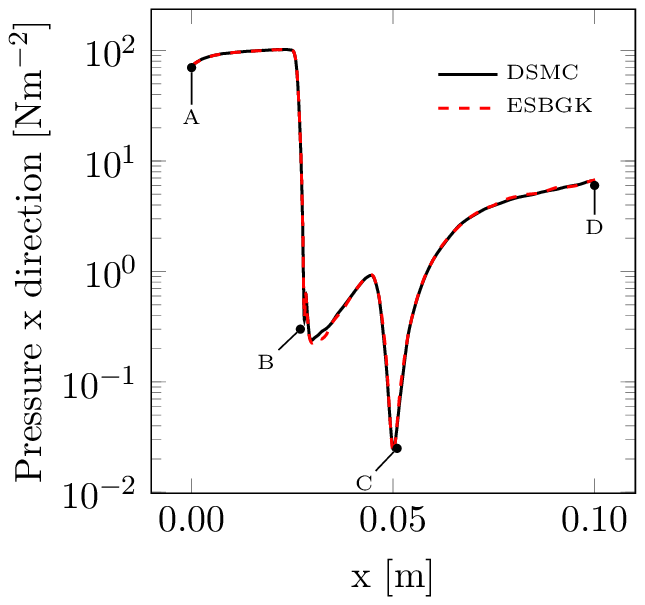}}
\caption{Simulation results of Case 2.}
\end{figure}

The temperature shock profile over the black line depicted in Fig. \ref{fig:shockline} is shown in detail in Fig. \ref{fig:shockcase2}. The overall agreement of the temperatures
is very good, whereby the largest difference is in the inflow area as previously discussed.

\begin{figure}
\centering
\includegraphics{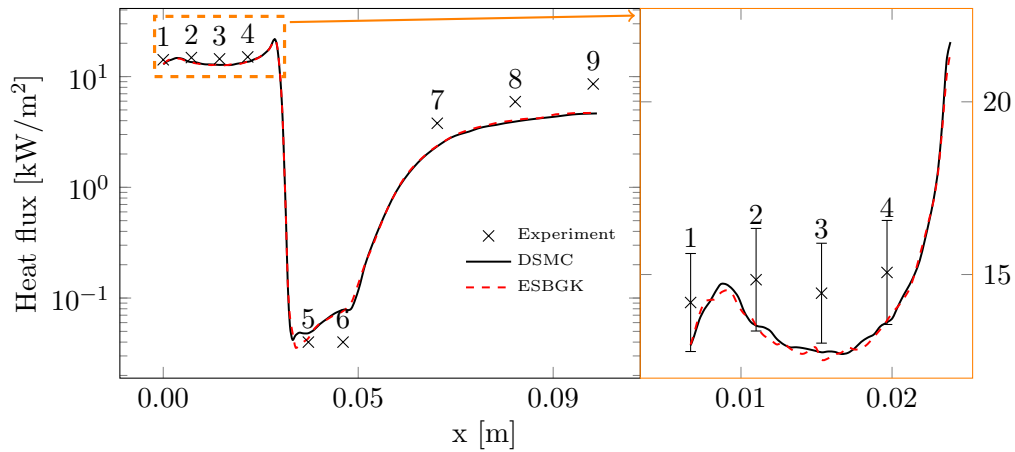}
\caption{Comparison of Heat flux on surface for Case 2.\label{fig:set1heatforceCase2}}
\end{figure}

The comparison of the heat flux and pressure in x-direction between DSMC and ESBGK are shown in Fig. \ref{fig:set1heatforceCase2} and \ref{fig:forcecase2}. 
The DSMC and ESBGK reults as well as the measurements show again a very well agreement.

A comparison of the computational time is shown in Table \ref{tab:cputimeset2}. 
The ESBGK method needs less particles $N_{ESBGK}=N_{DSMC}/8$ and allows a larger time step $t_{ESBGK}=2t_{DSMC}$. Therefore, the ESBGK model reduces the CPU time by a factor of $\approx 13.4$ 
for case 2 compared with DSMC. In this case, more particles are needed compared to case 1 to correctly resolve the temperature gradients.
Due to the angle of attack, a stronger bow shock forms in the front of the shield.
\begin{table}
  \begin{center}
\def~{\hphantom{0}}
  \begin{tabular}{l | C{2cm} C{2cm} C{3cm} C{4cm}}
   & Particle Number $N$ &  Time step $\Delta t$ [s] &  CPU Time / 100 iterations [s] & CPU Time / $1\cdot 10^{-5}\,\mathrm{s}$ Simulation time [s] \\
  \hline
  DSMC & $4.2\cdot 10^8$& $5\cdot 10^{-8}\,\mathrm{s}$  & 1636 & 3272 \\
  ESBGK & $N_{DSMC}/8$ & $2\Delta t_{DSMC}$ & 245 & 245
  \end{tabular}
  \caption{\label{tab:cputimeset2} Comparison of CPU time between DSMC and ESBGK for Case 2. The CPU time is the time per node with 24 cores on a Intel Xeon CPU E5-2680 v3.}
  \end{center}
\end{table}

\section{Conclusion}
A method is described that allows the efficient treatment of internal energies of diatomic molecules in the context of the particle-based ESBGK method. The shown method 
allows the handling of quantized as well as continuous vibrational energies. This allows the simulation of non-equilibrium low Knudsen number flows including diatomic molecules in a very
efficient way compared with DSMC simulations.

The method was verified using an adiabatic box with a non-equilibrium initial condition. It was shown that the ESBGK method is able to match the analytical temporal behaviour described
by the Landau-Teller equation as well as the DSMC results.

Further on, the ESBGK model was compared with DSMC simulations based on the hypersonic flow around a 70$^\circ$ blunted cone to 
evaluate the capabilities to capture the non-equilibrium effects in shock waves. It was shown that the heat flux values on the shield as well as the shock profiles fit
the DSMC results very well. Furthermore it was shown that the ESBGK method can save up to a factor of $\approx 35.8$ CPU time compared with DSMC for these problems.

This behavior is also very interesting for gas flows that cover a wide range of Knudsen numbers including continuum and rarefied gas regions as in nozzle expansion flows,
where the coupling of the proposed ESBGK method with DSMC is beneficial in order to save computational time.
The fact that DSMC and the investigated methods are both cell local Monte-Carlo based particle methods, makes a coupling very simple without the typical problems
of hybrid CFD-DSMC methods.

A next step will be the extension of the proposed method to gas mixtures to allow the simulation of more complex flows. 

\section*{Acknowledgments}

The author gratefully acknowledges the Deutsche Forschungsgemeinschaft (DFG) for funding this research within the
project “Partikelverfahren mit Strahlungslöser zur Simulation hochenthalper Nichtgleichgewichts-Plasmen” (project number 93159129). 
The author also thanks the High Performance Computing Center Stuttgart (HLRS) for granting the computational time that has allowed the execution of
the presented simulations.
 
\bibliography{mybibfile}

\begin{thebibliography}{23}%
\makeatletter
\providecommand \@ifxundefined [1]{%
 \@ifx{#1\undefined}
}%
\providecommand \@ifnum [1]{%
 \ifnum #1\expandafter \@firstoftwo
 \else \expandafter \@secondoftwo
 \fi
}%
\providecommand \@ifx [1]{%
 \ifx #1\expandafter \@firstoftwo
 \else \expandafter \@secondoftwo
 \fi
}%
\providecommand \natexlab [1]{#1}%
\providecommand \enquote  [1]{``#1''}%
\providecommand \bibnamefont  [1]{#1}%
\providecommand \bibfnamefont [1]{#1}%
\providecommand \citenamefont [1]{#1}%
\providecommand \href@noop [0]{\@secondoftwo}%
\providecommand \href [0]{\begingroup \@sanitize@url \@href}%
\providecommand \@href[1]{\@@startlink{#1}\@@href}%
\providecommand \@@href[1]{\endgroup#1\@@endlink}%
\providecommand \@sanitize@url [0]{\catcode `\\12\catcode `\$12\catcode
  `\&12\catcode `\#12\catcode `\^12\catcode `\_12\catcode `\%12\relax}%
\providecommand \@@startlink[1]{}%
\providecommand \@@endlink[0]{}%
\providecommand \url  [0]{\begingroup\@sanitize@url \@url }%
\providecommand \@url [1]{\endgroup\@href {#1}{\urlprefix }}%
\providecommand \urlprefix  [0]{URL }%
\providecommand \Eprint [0]{\href }%
\providecommand \doibase [0]{http://dx.doi.org/}%
\providecommand \selectlanguage [0]{\@gobble}%
\providecommand \bibinfo  [0]{\@secondoftwo}%
\providecommand \bibfield  [0]{\@secondoftwo}%
\providecommand \translation [1]{[#1]}%
\providecommand \BibitemOpen [0]{}%
\providecommand \bibitemStop [0]{}%
\providecommand \bibitemNoStop [0]{.\EOS\space}%
\providecommand \EOS [0]{\spacefactor3000\relax}%
\providecommand \BibitemShut  [1]{\csname bibitem#1\endcsname}%
\let\auto@bib@innerbib\@empty
\bibitem [{\citenamefont {Bird}(1994)}]{Bird1994}%
  \BibitemOpen
  \bibfield  {author} {\bibinfo {author} {\bibfnamefont {G.~A.}\ \bibnamefont
  {Bird}},\ }\href@noop {} {\emph {\bibinfo {title} {{Molecular Gas Dynamics
  and the Direct Simulation of Gas Flows}}}},\ \bibinfo {edition} {2nd}\ ed.\
  (\bibinfo  {publisher} {Oxford University Press},\ \bibinfo {address} {New
  York},\ \bibinfo {year} {1994})\BibitemShut {NoStop}%
\bibitem [{\citenamefont {Mirza}\ \emph {et~al.}(2017)\citenamefont {Mirza},
  \citenamefont {Nizenkov}, \citenamefont {Pfeiffer},\ and\ \citenamefont
  {Fasoulas}}]{MIRZA2017269}%
  \BibitemOpen
  \bibfield  {author} {\bibinfo {author} {\bibfnamefont {A.}~\bibnamefont
  {Mirza}}, \bibinfo {author} {\bibfnamefont {P.}~\bibnamefont {Nizenkov}},
  \bibinfo {author} {\bibfnamefont {M.}~\bibnamefont {Pfeiffer}}, \ and\
  \bibinfo {author} {\bibfnamefont {S.}~\bibnamefont {Fasoulas}},\ }\bibfield
  {title} {\enquote {\bibinfo {title} {{Three-dimensional implementation of the
  Low Diffusion method for continuum flow simulations}},}\ }\href {\doibase
  https://doi.org/10.1016/j.cpc.2017.07.018} {\bibfield  {journal} {\bibinfo
  {journal} {Computer Physics Communications}\ }\textbf {\bibinfo {volume}
  {220}},\ \bibinfo {pages} {269 -- 278} (\bibinfo {year} {2017})}\BibitemShut
  {NoStop}%
\bibitem [{\citenamefont {Burt}\ and\ \citenamefont
  {Boyd}(2006)}]{burt2006evaluation}%
  \BibitemOpen
  \bibfield  {author} {\bibinfo {author} {\bibfnamefont {J.}~\bibnamefont
  {Burt}}\ and\ \bibinfo {author} {\bibfnamefont {I.}~\bibnamefont {Boyd}},\
  }\bibfield  {title} {\enquote {\bibinfo {title} {{Evaluation of a particle
  method for the ellipsoidal statistical Bhatnagar-Gross-Krook equation}},}\
  }in\ \href@noop {} {\emph {\bibinfo {booktitle} {44th AIAA Aerospace Sciences
  Meeting and Exhibit}}}\ (\bibinfo {year} {2006})\ p.\ \bibinfo {pages}
  {989}\BibitemShut {NoStop}%
\bibitem [{\citenamefont {Titov}\ \emph {et~al.}(2008)\citenamefont {Titov},
  \citenamefont {Kumar}, \citenamefont {Levin}, \citenamefont {Gimelshein},\
  and\ \citenamefont {Gimelshein}}]{titov2008analysis}%
  \BibitemOpen
  \bibfield  {author} {\bibinfo {author} {\bibfnamefont {E.}~\bibnamefont
  {Titov}}, \bibinfo {author} {\bibfnamefont {R.}~\bibnamefont {Kumar}},
  \bibinfo {author} {\bibfnamefont {D.}~\bibnamefont {Levin}}, \bibinfo
  {author} {\bibfnamefont {N.}~\bibnamefont {Gimelshein}}, \ and\ \bibinfo
  {author} {\bibfnamefont {S.}~\bibnamefont {Gimelshein}},\ }\bibfield  {title}
  {\enquote {\bibinfo {title} {Analysis of different approaches to modeling of
  nozzle flows in the near continuum regime},}\ }in\ \href@noop {} {\emph
  {\bibinfo {booktitle} {AIP Conference Proceedings}}},\ Vol.\ \bibinfo
  {volume} {1084}\ (\bibinfo {organization} {AIP},\ \bibinfo {year} {2008})\
  pp.\ \bibinfo {pages} {978--984}\BibitemShut {NoStop}%
\bibitem [{\citenamefont {Tumuklu}, \citenamefont {Li},\ and\ \citenamefont
  {Levin}(2016)}]{tumuklu2016particle}%
  \BibitemOpen
  \bibfield  {author} {\bibinfo {author} {\bibfnamefont {O.}~\bibnamefont
  {Tumuklu}}, \bibinfo {author} {\bibfnamefont {Z.}~\bibnamefont {Li}}, \ and\
  \bibinfo {author} {\bibfnamefont {D.~A.}\ \bibnamefont {Levin}},\ }\bibfield
  {title} {\enquote {\bibinfo {title} {{Particle ellipsoidal statistical
  Bhatnagar-Gross-Krook approach for simulation of hypersonic shocks}},}\
  }\href@noop {} {\bibfield  {journal} {\bibinfo  {journal} {AIAA Journal}\ ,\
  \bibinfo {pages} {3701--3716}} (\bibinfo {year} {2016})}\BibitemShut
  {NoStop}%
\bibitem [{\citenamefont {Pfeiffer}()}]{Pfeiffer2018}%
  \BibitemOpen
  \bibfield  {author} {\bibinfo {author} {\bibfnamefont {M.}~\bibnamefont
  {Pfeiffer}},\ }\href@noop {} {\enquote {\bibinfo {title} {Particle-based
  fluid dynamics: comparison of different bhatnagar-gross-krook models and the
  direct simulation monte carlo method for hypersonic flows},}\ }\bibinfo
  {note} {Submitted}\BibitemShut {NoStop}%
\bibitem [{\citenamefont {Shakhov}(1968)}]{shakhov1968generalization}%
  \BibitemOpen
  \bibfield  {author} {\bibinfo {author} {\bibfnamefont {E.}~\bibnamefont
  {Shakhov}},\ }\bibfield  {title} {\enquote {\bibinfo {title} {{Generalization
  of the Krook kinetic relaxation equation}},}\ }\href@noop {} {\bibfield
  {journal} {\bibinfo  {journal} {Fluid Dynamics}\ }\textbf {\bibinfo {volume}
  {3}},\ \bibinfo {pages} {95--96} (\bibinfo {year} {1968})}\BibitemShut
  {NoStop}%
\bibitem [{\citenamefont {Holway~Jr}(1966)}]{holway1966new}%
  \BibitemOpen
  \bibfield  {author} {\bibinfo {author} {\bibfnamefont {L.~H.}\ \bibnamefont
  {Holway~Jr}},\ }\bibfield  {title} {\enquote {\bibinfo {title} {New
  statistical models for kinetic theory: methods of construction},}\
  }\href@noop {} {\bibfield  {journal} {\bibinfo  {journal} {The Physics of
  Fluids}\ }\textbf {\bibinfo {volume} {9}},\ \bibinfo {pages} {1658--1673}
  (\bibinfo {year} {1966})}\BibitemShut {NoStop}%
\bibitem [{\citenamefont {Gallis}\ and\ \citenamefont
  {Torczynski}(2011)}]{gallis2011investigation}%
  \BibitemOpen
  \bibfield  {author} {\bibinfo {author} {\bibfnamefont {M.}~\bibnamefont
  {Gallis}}\ and\ \bibinfo {author} {\bibfnamefont {J.}~\bibnamefont
  {Torczynski}},\ }\bibfield  {title} {\enquote {\bibinfo {title}
  {{Investigation of the ellipsoidal-statistical Bhatnagar-Gross-Krook kinetic
  model applied to gas-phase transport of heat and tangential momentum between
  parallel walls}},}\ }\href@noop {} {\bibfield  {journal} {\bibinfo  {journal}
  {Physics of Fluids}\ }\textbf {\bibinfo {volume} {23}},\ \bibinfo {pages}
  {030601} (\bibinfo {year} {2011})}\BibitemShut {NoStop}%
\bibitem [{\citenamefont {Bhatnagar}, \citenamefont {Gross},\ and\
  \citenamefont {Krook}(1954)}]{bhatnagar1954model}%
  \BibitemOpen
  \bibfield  {author} {\bibinfo {author} {\bibfnamefont {P.~L.}\ \bibnamefont
  {Bhatnagar}}, \bibinfo {author} {\bibfnamefont {E.~P.}\ \bibnamefont
  {Gross}}, \ and\ \bibinfo {author} {\bibfnamefont {M.}~\bibnamefont
  {Krook}},\ }\bibfield  {title} {\enquote {\bibinfo {title} {{A model for
  collision processes in gases. I. Small amplitude processes in charged and
  neutral one-component systems}},}\ }\href@noop {} {\bibfield  {journal}
  {\bibinfo  {journal} {Physical review}\ }\textbf {\bibinfo {volume} {94}},\
  \bibinfo {pages} {511} (\bibinfo {year} {1954})}\BibitemShut {NoStop}%
\bibitem [{\citenamefont {Andries}\ \emph {et~al.}(2000)\citenamefont
  {Andries}, \citenamefont {Le~Tallec}, \citenamefont {Perlat},\ and\
  \citenamefont {Perthame}}]{andries2000gaussian}%
  \BibitemOpen
  \bibfield  {author} {\bibinfo {author} {\bibfnamefont {P.}~\bibnamefont
  {Andries}}, \bibinfo {author} {\bibfnamefont {P.}~\bibnamefont {Le~Tallec}},
  \bibinfo {author} {\bibfnamefont {J.-P.}\ \bibnamefont {Perlat}}, \ and\
  \bibinfo {author} {\bibfnamefont {B.}~\bibnamefont {Perthame}},\ }\bibfield
  {title} {\enquote {\bibinfo {title} {{The Gaussian-BGK model of Boltzmann
  equation with small Prandtl number}},}\ }\href@noop {} {\bibfield  {journal}
  {\bibinfo  {journal} {European Journal of Mechanics-B/Fluids}\ }\textbf
  {\bibinfo {volume} {19}},\ \bibinfo {pages} {813--830} (\bibinfo {year}
  {2000})}\BibitemShut {NoStop}%
\bibitem [{\citenamefont {Andries}\ and\ \citenamefont
  {Perthame}(2001)}]{andries2001bgk}%
  \BibitemOpen
  \bibfield  {author} {\bibinfo {author} {\bibfnamefont {P.}~\bibnamefont
  {Andries}}\ and\ \bibinfo {author} {\bibfnamefont {B.}~\bibnamefont
  {Perthame}},\ }\bibfield  {title} {\enquote {\bibinfo {title} {{The ES-BGK
  model equation with correct Prandtl number}},}\ }in\ \href@noop {} {\emph
  {\bibinfo {booktitle} {AIP conference proceedings}}},\ Vol.\ \bibinfo
  {volume} {585}\ (\bibinfo {organization} {AIP},\ \bibinfo {year} {2001})\
  pp.\ \bibinfo {pages} {30--36}\BibitemShut {NoStop}%
\bibitem [{\citenamefont {Zhang}\ and\ \citenamefont
  {Schwartzentruber}(2013)}]{zhang:2013}%
  \BibitemOpen
  \bibfield  {author} {\bibinfo {author} {\bibfnamefont {C.}~\bibnamefont
  {Zhang}}\ and\ \bibinfo {author} {\bibfnamefont {T.~E.}\ \bibnamefont
  {Schwartzentruber}},\ }\bibfield  {title} {\enquote {\bibinfo {title}
  {Inelastic collision selection procedures for direct simulation monte carlo
  calculations of gas mixtures},}\ }\href {\doibase 10.1063/1.4825340}
  {\bibfield  {journal} {\bibinfo  {journal} {Physics of Fluids
  (1994-present)}\ }\textbf {\bibinfo {volume} {25}},\ \bibinfo {eid} {106105}
  (\bibinfo {year} {2013})}\BibitemShut {NoStop}%
\bibitem [{\citenamefont {Gimelshein}, \citenamefont {Gimelshein},\ and\
  \citenamefont {Levin}(2002)}]{Gimelshein2002}%
  \BibitemOpen
  \bibfield  {author} {\bibinfo {author} {\bibfnamefont {N.~E.}\ \bibnamefont
  {Gimelshein}}, \bibinfo {author} {\bibfnamefont {S.~F.}\ \bibnamefont
  {Gimelshein}}, \ and\ \bibinfo {author} {\bibfnamefont {D.~A.}\ \bibnamefont
  {Levin}},\ }\bibfield  {title} {\enquote {\bibinfo {title} {{Vibrational
  relaxation rates in the direct simulation Monte Carlo method}},}\ }\href
  {\doibase 10.1063/1.1517297} {\bibfield  {journal} {\bibinfo  {journal}
  {Physics of Fluids}\ }\textbf {\bibinfo {volume} {14}},\ \bibinfo {pages}
  {4452} (\bibinfo {year} {2002})}\BibitemShut {NoStop}%
\bibitem [{\citenamefont {Pfeiffer}\ \emph
  {et~al.}(2016{\natexlab{a}})\citenamefont {Pfeiffer}, \citenamefont
  {Nizenkov}, \citenamefont {Mirza},\ and\ \citenamefont
  {Fasoulas}}]{pfeiffer2016direct}%
  \BibitemOpen
  \bibfield  {author} {\bibinfo {author} {\bibfnamefont {M.}~\bibnamefont
  {Pfeiffer}}, \bibinfo {author} {\bibfnamefont {P.}~\bibnamefont {Nizenkov}},
  \bibinfo {author} {\bibfnamefont {A.}~\bibnamefont {Mirza}}, \ and\ \bibinfo
  {author} {\bibfnamefont {S.}~\bibnamefont {Fasoulas}},\ }\bibfield  {title}
  {\enquote {\bibinfo {title} {{Direct Simulation Monte Carlo modeling of
  relaxation processes in polyatomic gases}},}\ }\href@noop {} {\bibfield
  {journal} {\bibinfo  {journal} {Physics of Fluids}\ }\textbf {\bibinfo
  {volume} {28}},\ \bibinfo {pages} {027103} (\bibinfo {year}
  {2016}{\natexlab{a}})}\BibitemShut {NoStop}%
\bibitem [{\citenamefont {Sun}\ and\ \citenamefont
  {Boyd}(2005)}]{sun2005evaluation}%
  \BibitemOpen
  \bibfield  {author} {\bibinfo {author} {\bibfnamefont {Q.}~\bibnamefont
  {Sun}}\ and\ \bibinfo {author} {\bibfnamefont {I.~D.}\ \bibnamefont {Boyd}},\
  }\bibfield  {title} {\enquote {\bibinfo {title} {{Evaluation of macroscopic
  properties in the direct simulation Monte Carlo method}},}\ }\href@noop {}
  {\bibfield  {journal} {\bibinfo  {journal} {Journal of Thermophysics and Heat
  Transfer}\ }\textbf {\bibinfo {volume} {19}},\ \bibinfo {pages} {329--335}
  (\bibinfo {year} {2005})}\BibitemShut {NoStop}%
\bibitem [{\citenamefont {Haas}\ \emph {et~al.}(1994)\citenamefont {Haas},
  \citenamefont {Hash}, \citenamefont {Bird}, \citenamefont {Lumpkin},\ and\
  \citenamefont {Hassan}}]{Haas1994}%
  \BibitemOpen
  \bibfield  {author} {\bibinfo {author} {\bibfnamefont {B.~L.}\ \bibnamefont
  {Haas}}, \bibinfo {author} {\bibfnamefont {D.~B.}\ \bibnamefont {Hash}},
  \bibinfo {author} {\bibfnamefont {G.~A.}\ \bibnamefont {Bird}}, \bibinfo
  {author} {\bibfnamefont {F.~E.}\ \bibnamefont {Lumpkin}}, \ and\ \bibinfo
  {author} {\bibfnamefont {H.~A.}\ \bibnamefont {Hassan}},\ }\bibfield  {title}
  {\enquote {\bibinfo {title} {{Rates of thermal relaxation in direct
  simulation Monte Carlo methods}},}\ }\href {\doibase 10.1063/1.868221}
  {\bibfield  {journal} {\bibinfo  {journal} {Physics of Fluids}\ }\textbf
  {\bibinfo {volume} {6}},\ \bibinfo {pages} {2191} (\bibinfo {year}
  {1994})}\BibitemShut {NoStop}%
\bibitem [{\citenamefont {Munz}\ \emph {et~al.}(2014)\citenamefont {Munz},
  \citenamefont {Auweter-Kurtz}, \citenamefont {Fasoulas}, \citenamefont
  {Mirza}, \citenamefont {Ortwein}, \citenamefont {Pfeiffer},\ and\
  \citenamefont {Stindl}}]{Munz2014}%
  \BibitemOpen
  \bibfield  {author} {\bibinfo {author} {\bibfnamefont {C.-D.}\ \bibnamefont
  {Munz}}, \bibinfo {author} {\bibfnamefont {M.}~\bibnamefont {Auweter-Kurtz}},
  \bibinfo {author} {\bibfnamefont {S.}~\bibnamefont {Fasoulas}}, \bibinfo
  {author} {\bibfnamefont {A.}~\bibnamefont {Mirza}}, \bibinfo {author}
  {\bibfnamefont {P.}~\bibnamefont {Ortwein}}, \bibinfo {author} {\bibfnamefont
  {M.}~\bibnamefont {Pfeiffer}}, \ and\ \bibinfo {author} {\bibfnamefont
  {T.}~\bibnamefont {Stindl}},\ }\bibfield  {title} {\enquote {\bibinfo {title}
  {{Coupled Particle-In-Cell and Direct Simulation Monte Carlo method for
  simulating reactive plasma flows}},}\ }\href {\doibase
  10.1016/j.crme.2014.07.005} {\bibfield  {journal} {\bibinfo  {journal}
  {Comptes Rendus M{\'{e}}canique}\ }\textbf {\bibinfo {volume} {342}},\
  \bibinfo {pages} {662--670} (\bibinfo {year} {2014})}\BibitemShut {NoStop}%
\bibitem [{\citenamefont {Gallis}\ and\ \citenamefont
  {Torczynski}(2000)}]{gallis2000application}%
  \BibitemOpen
  \bibfield  {author} {\bibinfo {author} {\bibfnamefont {M.}~\bibnamefont
  {Gallis}}\ and\ \bibinfo {author} {\bibfnamefont {J.}~\bibnamefont
  {Torczynski}},\ }\bibfield  {title} {\enquote {\bibinfo {title} {{The
  application of the BGK model in particle simulations}},}\ }in\ \href@noop {}
  {\emph {\bibinfo {booktitle} {34th Thermophysics Conference}}}\ (\bibinfo
  {year} {2000})\ p.\ \bibinfo {pages} {2360}\BibitemShut {NoStop}%
\bibitem [{\citenamefont {Pfeiffer}\ \emph
  {et~al.}(2016{\natexlab{b}})\citenamefont {Pfeiffer}, \citenamefont
  {Nizenkov}, \citenamefont {Mirza},\ and\ \citenamefont
  {Fasoulas}}]{Pfeiffer2016b}%
  \BibitemOpen
  \bibfield  {author} {\bibinfo {author} {\bibfnamefont {M.}~\bibnamefont
  {Pfeiffer}}, \bibinfo {author} {\bibfnamefont {P.}~\bibnamefont {Nizenkov}},
  \bibinfo {author} {\bibfnamefont {A.}~\bibnamefont {Mirza}}, \ and\ \bibinfo
  {author} {\bibfnamefont {S.}~\bibnamefont {Fasoulas}},\ }\bibfield  {title}
  {\enquote {\bibinfo {title} {{Direct simulation Monte Carlo modeling of
  relaxation processes in polyatomic gases}},}\ }\href {\doibase
  10.1063/1.4940989} {\bibfield  {journal} {\bibinfo  {journal} {Physics of
  Fluids}\ }\textbf {\bibinfo {volume} {28}},\ \bibinfo {pages} {027103}
  (\bibinfo {year} {2016}{\natexlab{b}})}\BibitemShut {NoStop}%
\bibitem [{\citenamefont {All{\`{e}}gre}, \citenamefont {Bisch},\ and\
  \citenamefont {Lengrand}(1997)}]{Allegre1997}%
  \BibitemOpen
  \bibfield  {author} {\bibinfo {author} {\bibfnamefont {J.}~\bibnamefont
  {All{\`{e}}gre}}, \bibinfo {author} {\bibfnamefont {D.}~\bibnamefont
  {Bisch}}, \ and\ \bibinfo {author} {\bibfnamefont {J.~C.}\ \bibnamefont
  {Lengrand}},\ }\bibfield  {title} {\enquote {\bibinfo {title} {{Experimental
  Rarefied Heat Transfer at Hypersonic Conditions over 70-Degree Blunted
  Cone}},}\ }\href {\doibase 10.2514/2.3302} {\bibfield  {journal} {\bibinfo
  {journal} {Journal of Spacecraft and Rockets}\ }\textbf {\bibinfo {volume}
  {34}},\ \bibinfo {pages} {724--728} (\bibinfo {year} {1997})}\BibitemShut
  {NoStop}%
\bibitem [{\citenamefont {Nizenkov}\ \emph {et~al.}(2017)\citenamefont
  {Nizenkov}, \citenamefont {Noeding}, \citenamefont {Konopka},\ and\
  \citenamefont {Fasoulas}}]{Nizenkov2017}%
  \BibitemOpen
  \bibfield  {author} {\bibinfo {author} {\bibfnamefont {P.}~\bibnamefont
  {Nizenkov}}, \bibinfo {author} {\bibfnamefont {P.}~\bibnamefont {Noeding}},
  \bibinfo {author} {\bibfnamefont {M.}~\bibnamefont {Konopka}}, \ and\
  \bibinfo {author} {\bibfnamefont {S.}~\bibnamefont {Fasoulas}},\ }\bibfield
  {title} {\enquote {\bibinfo {title} {Verification and validation of a
  parallel 3d direct simulation monte carlo solver for atmospheric entry
  applications},}\ }\href {\doibase 10.1007/s12567-016-0133-5} {\bibfield
  {journal} {\bibinfo  {journal} {CEAS Space Journal}\ }\textbf {\bibinfo
  {volume} {9}},\ \bibinfo {pages} {127--137} (\bibinfo {year}
  {2017})}\BibitemShut {NoStop}%
\bibitem [{\citenamefont {Pfeiffer}\ and\ \citenamefont
  {Gorji}(2017)}]{PFEIFFER20171}%
  \BibitemOpen
  \bibfield  {author} {\bibinfo {author} {\bibfnamefont {M.}~\bibnamefont
  {Pfeiffer}}\ and\ \bibinfo {author} {\bibfnamefont {M.}~\bibnamefont
  {Gorji}},\ }\bibfield  {title} {\enquote {\bibinfo {title} {{Adaptive
  Particle-Cell algorithm for Fokker-Planck based rarefied gas flow
  simulations}},}\ }\href {\doibase https://doi.org/10.1016/j.cpc.2016.11.003}
  {\bibfield  {journal} {\bibinfo  {journal} {Computer Physics Communications}\
  }\textbf {\bibinfo {volume} {213}},\ \bibinfo {pages} {1 -- 8} (\bibinfo
  {year} {2017})}\BibitemShut {NoStop}%
\end{thebibliography}%
\end{document}